\documentclass[10pt,letterpaper]{article}

\usepackage{changepage}

\usepackage[utf8]{inputenc}

\usepackage{textcomp,marvosym}

\usepackage{fixltx2e}

\usepackage{amsmath,amssymb}

\usepackage{cite}

\usepackage{nameref,hyperref}

\usepackage[right]{lineno}

\usepackage{microtype}
\DisableLigatures[f]{encoding = *, family = * }

\usepackage{rotating}

\usepackage{soul}

\date{}

\newcommand{\acc}{$\mathit{Accuracy}$}
\newcommand{\accone}{$\mathit{Accuracy}\!\pm\!1$}
\newcommand{\prc}{$\mathit{Precision}$}
\newcommand{\rcl}{$\mathit{Recall}$}

\usepackage{color}

\newcommand{\sn}{\hphantom{*}\hphantom{*}}
\newcommand{\se}{*\hphantom{*}}
\newcommand{\sd}{**}

\begin{document}

\vspace*{0.35in}

\begin{flushleft}
{\Large
\textbf\newline{The Effects of Twitter Sentiment on Stock Price Returns}
}
\newline
\\
Gabriele Ranco\textsuperscript{1},
Darko Aleksovski\textsuperscript{2,*}, 
Guido Caldarelli\textsuperscript{1,3,4},
Miha Gr\v{c}ar\textsuperscript{2},
Igor Mozeti\v{c}\textsuperscript{2}
\\
\bf{1} IMT Institute for Advanced Studies, Piazza San Francesco 19, 55100 Lucca, Italy
\\
\bf{2} Jo\v{z}ef Stefan Institute, Jamova 39, 1000 Ljubljana, Slovenia
\\
\bf{3} Istituto dei Sistemi Complessi (ISC), Via dei Taurini 19, 00185 Rome, Italy
\\
\bf{4} London Institute for Mathematical Sciences, 35a South St. Mayfair, London W1K 2XF, UK
\\

* darko.aleksovski@ijs.si

\end{flushleft}


\section*{Abstract}
Social media are increasingly reflecting and influencing behavior of other
complex systems. In this paper we investigate the relations between a 
well-know micro-blogging platform Twitter and financial markets. 
In particular, we consider, in a period of 15 months, the Twitter volume 
and sentiment about the 30 stock companies
that form the Dow Jones Industrial Average (DJIA) index.
We find a relatively low Pearson correlation and Granger causality 
between the corresponding time series over the entire time period. 
However, we find a significant dependence between the Twitter sentiment and 
abnormal returns during the peaks of Twitter volume. 
This is valid not only for the expected Twitter volume 
peaks (e.g., quarterly announcements), but also for peaks corresponding 
to less obvious events.
We formalize the procedure by adapting the well-known
``event study'' from economics and finance to the analysis of Twitter data.
The procedure allows to automatically identify events as 
Twitter volume peaks, to compute the prevailing
sentiment (positive or negative) expressed in tweets at these peaks,
and finally to apply the ``event study'' methodology to relate them to stock returns.
We show that sentiment polarity of Twitter peaks implies
the direction of cumulative abnormal returns. The amount of cumulative
abnormal returns is relatively low (about 1--2\%), but the dependence is
statistically significant for several days after the events. 

\section*{Introduction}
The recent technological revolution with widespread presence of computers and Internet
has created an unprecedented situation of data deluge, changing dramatically the way in which we look
at social and economic sciences. The constantly increasing use of the Internet as a source of information, such as business or political news, triggered an analogous increasing online activity.
The interaction with technological systems is generating massive datasets that document collective behavior in a previously unimaginable fashion \cite{king2011ensuring,vespignani2009predicting}.
Ultimately, in this vast repository of Internet activity we can find the interests, concerns, and intentions of the global population with respect to various economic, political, and cultural phenomena.

Among the many fields of applications of data collection, analysis and modeling, we present here a case study on financial systems.
We believe that social aspects as measured by social networks are particularly useful to understand financial turnovers. Indeed, financial contagion and, ultimately, crises, are often originated by collective phenomena such
as herding among investors (or, in extreme cases, panic) which signal the intrinsic complexity of the
financial system \cite{bouchaud2009unfortunate}. Therefore, the possibility to anticipate anomalous collective behavior of investors is
of great interest to policy makers \cite{haldane2011systemic, fagiolo2009economic, bouchaud2008economics} because it may allow for a more prompt intervention, when appropriate.

\paragraph*{State-of-the-art.}

We briefly review the state-of-the-art research which investigates the
correlation between the web data and financial markets.
Three major classes of data are considered: web news, search engine queries,
and social media.
Regarding news, various approaches have been attempted. They study:
(i)  the connection of exogenous news with price movements\cite{cutler1998moves},
(ii)  the stock price reaction to news\cite{chan2003stock,vega2006stock};
(iii) the relations between mentions of a company in financial news \cite{alanyali2013quantifying}, or the pessimism of the media\cite{tetlock2007giving}, and trading volume;
(iv)  the relation between the sentiment of news, earnings and return predictability\cite{tetlock2008more},
(v)  the role of news in  trading actions \cite{lillo2012how}, 
especially of short sellers \cite{engelberg2012shorts};
(vi) the role of macroeconomic news in stock returns \cite{birz2011effect};
and finally (vii) the high-frequency market reactions to news \cite{gross2011machines}.

There are several analyses of search engine queries.
A relation between the daily number of queries for a particular stock, and daily 
trading volume of the same stock has been studied by 
\cite{preis2010complex,bordino2012web,bordino2014stock}.
A similar analysis was done for a sample of Russell 3000 stocks, 
where an increase in queries predicts higher stock prices in the next two weeks \cite{da2011search}.
Search engine query data from Google Trends has been used to evaluate stock riskiness \cite{kristoufek2013can}.
Some other authors used Google trends to predict market
movements \cite{curme2014quantifying}.
Also, search engine query data has been used as a proxy for analyzing investor attention related to initial public offerings (IPOs) \cite{vakrman2015underpricing}.

Regarding social media, Twitter is becoming an increasingly popular micro-blogging platform used
for financial forecasting \cite{Graham2013,nguyen2013royal,akshay2007why}.
One line of research investigates the relation between the volume of tweets
and financial markets.
For example, \cite{mao2012correlating} studied
whether the daily number of tweets predicts the S\&P 500 stock indicators.
Another line of research explores the contents of tweets.
In a textual analysis approach to Twitter data, 
the authors 
find clear relations between the 
mood indicators and Dow Jones Industrial Average (DJIA) \cite{bollen2011twitter,bollen2011modeling,mao2011predicting}.
In \cite{Souza2015}, the authors show that the Twitter sentiment for five retail
companies has statistically significant relation with stock returns and volatility.
A recent study \cite{Zheludev2014} compares the information content of the
Twitter sentiment and volume in terms of their influence on future stock prices.
The authors relate the intra-day Twitter and price data, at hourly resolution,
and show that the Twitter sentiment contains significantly more lead-time
information about the prices than the Twitter volume alone.
They apply stringent statistics which require relatively high volume of
tweets over the entire period of three months, and, as a consequence, only
12 financial instruments pass the test.

\paragraph*{Motivation.}
Despite the high quality of the data sets used, the level of 
empirical correlation between stock price derived financial time series 
and web derived time series remains limited, 
especially when a textual analysis of web messages is applied.
This observation suggests that the relation between these two systems is 
more complex and that a simple measure of  correlation is not 
enough to capture the dynamics of the interaction between the two systems. 
It is possible that the two systems are dependent only at some moments 
of their evolution, and not over the entire time period.

In this paper, we study the relation between stock price returns and
the sentiment expressed in financial tweets posted on Twitter.
We analyze a carefully collected and annotated set of tweets
about the previously-mentioned 30 DJIA companies.
For each of these companies we build a
time series of the sentiment expressed in the tweets, with  
daily resolution, designed to mimic 
the wisdom-of-crowd effect, as observed in previous works.
As first analysis we compute the Pearson correlation between price return 
time series and the sentiment time series generated from the tweets.
We also run a Granger causality test \cite{granger1969investigating} to study 
the forecasting power of the Twitter time series.
When considering the entire period of 15 months, 
the values of Pearson correlation are low 
and only a few companies pass the Granger causality test.

In order to detect the presence of a stronger correlation, at least 
in some portions of the  time series,
we consider the relation between the stock price returns and Twitter sentiment
through the technique of ``event study'' \cite{campbell1997econometrics,boehmer1991event},
known in economics and finance.
This  technique has been generally used to verify if the sentiment content of 
earnings announcements conveys useful information for the valuation of companies.
Here we apply a similar approach, but instead of using the sentiment of earnings 
announcement, we use the aggregate sentiment expressed in financial tweets.

\paragraph*{Contributions.}
By restricting our analysis to shorter time periods around the ``events'' we find a 
statistically significant relation between the Twitter sentiment and stock returns.
These results are consistent with the existing literature on
the information content of earnings \cite{campbell1997econometrics,boehmer1991event}.
A recent related study \cite{Sprenger2014jbfa,Sprenger2014efm},
also applies the ``event study'' methodology to Twitter data.
The authors come to similar conclusions as we do: financial Twitter data,
when considering both, the volume and sentiment of tweets, does have
a statistically significant impact on stock returns. 
It is interesting that two independent studies, to the best of our knowledge 
the first adaptations of ``event study'' to Twitter data, corroborate the conclusions.

This paper presents a complementary study to \cite{Sprenger2014jbfa}, and uses a slightly different experimental setup. The studies use disjoint sets of stocks (S\&P 500 vs. DJIA 30), non-overlapping time windows (January--June 2010 vs. June 2013--September 2014), different sentiment classification techniques (Naive Bayes vs. Support Vector Machine), different event detection algorithms, and different statistics for significance testing. We point out the differences between the two studies in the appropriate sections of the paper. Despite the methodological differences, we can confirm the main results reported in \cite{Sprenger2014jbfa} to a large extent. 
From this perspective, one of the contributions of this work is in providing even more evidence, over a longer time period, for the conclusions drawn in both studies.

The second contribution is that the Twitter sentiment time series are made publicly available. They can be used not only to validate our results but also to carry out additional studies that do not necessarily follow the same methodology. The dataset allows one to study different sentiment aggregations, different events (points in time), and different post-event effects (such as drifts, reversals, and changes in volatility rather than abnormal returns). 

The third contribution, as compared to \cite{Sprenger2014jbfa}, is the use of a high 
quality sentiment classifier, and the realistic evaluation of its performance.
Our sentiment classifier was trained on a much larger training set
(2,500 vs. over 100,000 annotated tweets in our case), and exhaustively evaluated.
This resulted in the performance that matches the agreement
between financial experts.
The human annotation of such large number of tweets is relatively expensive.
However, there are several advantages.
First, a considerable amount of tweets can be annotated twice, by two different
annotators, in order to compute the inter-annotator agreement and thus establish
an upper bound on the performance.
Second, there is no need to collect domain-specific vocabularies,
since the annotation process itself is domain and language specific.
Third, once a large enough set of tweets is assigned a sentiment label,
the classifier construction is automated and the domain-specific sentiment models
are available for real-time processing.

We have already applied the same sentiment classification methodology
in various domains, such as:
(i) to study the emotional dynamics of Facebook comments on conspiracy 
theories (in Italian) \cite{Zollo2015}, 
(ii) to compare the sentiment leaning of different network communities 
towards various environmental topics \cite{Sluban2015}, and
(iii) to monitor the sentiment about political parties before and after 
the elections (in Bulgarian) \cite{smailovic2015}.

\section*{Data}
\label{secdata4}
Our analysis is conducted on 30 stocks of the DJIA index. The stock data are collected 
for a period of 15 months between 2013 and 2014.
The ticker list of the investigated stocks is shown in Table \ref{tbl_stocks}. 
In the analysis we investigate the relation between price/market data, and Twitter data. The details of both are given in the remainder of this section.

\begin{table}
\centering
\begin{tabular}{llr}
\hline
Ticker &                        Company &  Tweets \\
\hline
TRV  &            Travelers Companies Corp &                   12,184 \\
UNH  &              UnitedHealth Group Inc &                   15,020 \\
UTX  &            United Technologies Corp &                   16,123 \\
MMM  &                               3M Co &                   17,001 \\
DD   &       E I du Pont de Nemours and Co &                   17,340 \\
AXP  &                 American Express Co &                   21,941 \\
PG   &                Procter \& Gamble Co &                   25,751 \\
NKE  &                            Nike Inc &                   29,220 \\
CVX  &                        Chevron Corp &                   29,477 \\
HD   &                      Home Depot Inc &                   30,923 \\
CAT  &                     Caterpillar Inc &                   38,739 \\
JNJ  &                  Johnson \& Johnson &                   40,503 \\
V    &                            Visa Inc &                   43,375 \\
VZ   &          Verizon Communications Inc &                   45,177 \\
KO   &                        Coca-Cola Co &                   45,339 \\
MCD  &                     McDonald's Corp &                   45,971 \\
XOM  &                    Exxon Mobil Corp &                   46,286 \\
DIS  &                      Walt Disney Co &                   46,439 \\
BA   &                           Boeing Co &                   51,799 \\
MRK  &                     Merck \& Co Inc &                   54,986 \\
CSCO &                   Cisco Systems Inc &                   57,427 \\
GE   &                 General Electric Co &                   61,836 \\
WMT  &                 Wal-Mart Stores Inc &                   63,405 \\
INTC &                          Intel Corp &                   68,079 \\
PFE  &                          Pfizer Inc &                   71,415 \\
T    &                           AT\&T Inc &                   75,886 \\
GS   &             Goldman Sachs Group Inc &                   91,057 \\
IBM  &  International Business Machines Co &                  101,077 \\
JPM  &               JPMorgan Chase and Co &                  108,810 \\
MSFT &                      Microsoft Corp &                  183,184 \\
\hline
Total   &                               &     1,555,770 \\
\hline
\end{tabular}
\caption{The collected Twitter data for the 15 months period:
the company names and the number of tweets.}
\label{tbl_stocks}

\end{table}

\subsection*{Market data}
The first source of data contains information on price returns of the stock, with daily resolution.
For each stock we extract the time series of daily returns, $R_d$:
\begin{equation}
R_d=\frac{p_{d}-p_{d-1}}{p_{d-1}}
\end{equation}
where $p_{d}$ is the closing price of the stock at day $d$.
We use raw-returns, and not the more standard log-returns, to be consistent
with the original ``event study'' \cite{campbell1997econometrics,mackinlay1997event}.
This data is publicly available and can be downloaded from various
sources on the Internet, as for example the Nasdaq web site \footnote{\url{http://www.nasdaq.com/symbol/nke/historical} for the ``Nike'' stock}.

\subsection*{Twitter data}
The second source of data is from Twitter and consists of relevant
tweets, along with their sentiment.
The data was collected by Twitter Search API, where a search query consists
of the stock cash-tag (e.g., ``\$NKE'' for Nike).
To the best of our knowledge, all the available tweets with cash-tags
are acquired. The Twitter restriction of 1\% (or 10\%) of tweets applies 
to the Twitter Streaming API, and only in the case when the specified filter 
(query) is general enough to account for more than 1\% (or 10\%) of all public tweets.
The data covers a period of 15 months (from June 1, 2013 to September 18, 2014),
for which there is 
over 1.5 million tweets.
The tweets 
for the analysis were provided to us 
by the Sowa Labs company (\url{http://www.sowalabs.com/}).

The Twitter sentiment is calculated by a supervised learning method.
First, over 100,000 of tweets were labeled 
by 10 financial experts with three sentiment labels: negative, neutral or positive. 
Then, this labeled set was used to build a 
Support Vector Machine (SVM \cite{vapnik95}) classification model
which discriminates between  negative, neutral and positive tweets.
Finally, the SVM model  was applied to the complete set of over 1.5 million tweets. 
The resulting data set 
is in the form of a time series of negative, neutral and positive tweets for each day $d$.
In particular, we create the following time series for each company:
\begin{itemize}
\item Volume of tweets, $TW_d$: the total number of tweets in a day. 
\item Negative tweets, $tw^{-}_{d}$: the number of negative tweets in a day.
\item Neutral tweets, $tw^{0}_{d}$: the number of neutral tweets in a day.
\item Positive tweets, $tw^{+}_{d}$: the number of positive tweets in a day.
\item Sentiment polarity, $P_d$: 
the difference between the number of positive and negative tweets as a
fraction of non-neutral tweets\cite{zhang2010trading}, 
$P_d=\frac{tw^{+}_{d}-tw^{-}_{d}}{tw^{+}_{d}+tw^{-}_{d}}$.
\end{itemize}

The Twitter sentiment and financial time series data for the DJIA 30 stocks are available at \url{http://kt.ijs.si/data/Twitter\_sentiment\_DJIA30/}.

\section*{Methods}
\label{sec:Meth4}

This section first describes the machine learning methodology used for 
sentiment classification. Then, it presents the methods used for 
the correlation analysis and Granger causality. 
Finally, it describes the event study methodology, by presenting the detection of events, 
the categorization of events based on Twitter sentiment, 
and the statistical validation of the cumulative abnormal returns.

\subsection*{Sentiment classification}

Determining sentiment polarity of tweets is not an easy task.
Financial experts often disagree whether a given tweet represents a buy or
a sell signal, and even individuals are not always consistent with themselves.
We argue that the upper bound that any automated sentiment classification
procedure can achieve is determined by the level of agreement between 
the human experts. 
In order to achieve the performance of human experts, a large enough
set of tweets has to be manually annotated -- in our case, over 100,000.
In order to measure the agreement between the experts,
a substantial fraction of tweets has to be annotated by two
different experts -- in our case, over 6,000 tweets were annotated twice.

Our approach to automatic sentiment classification of tweets is based
on supervised machine learning. The procedure consists of the following steps:
(i) a sample of tweets is manually annotated with sentiment,
(ii) the labeled set is used to train and tune a classifier,
(iii) the classifier is evaluated by cross-validation and
compared to the inter-annotator agreement, and
(iv) the classifier is applied to the whole set of collected tweets.

In this paper, as is common in the sentiment analysis literature \cite{pang08},
we have approximated the sentiment of tweets with an ordinal scale of three values:
\textit{negative} ($-$), \textit{neutral} ($0$), and \textit{positive} ($+$).
Sentiment classification is an ordinal classification task,
a special case of multi-class classification where 
there is a natural ordering between the classes, 
but no meaningful numeric difference between them \cite{gaudette2009evaluation}.
Our classifier is based on Support Vector Machine (SVM),
a widely used, state-of-the-art supervised learning algorithm,
well suited for large scale text categorization tasks,
and robust on large feature spaces. 
We implemented the wrapper approach, described in \cite{frank2001simple},
which constructs two linear-kernel SVM \cite{vapnik95} classifiers.
Since the classes are ordered, two classifiers suffice to partition the
space of tweets into the three sentiment areas.
The two SVM classifiers were trained to distinguish
between \textit{positive} and \textit{negative-or-neutral}, and between \textit{negative} and \textit{positive-or-neutral}, respectively.
During prediction, if the target class cannot be determined as the two classifiers 
disagree (which happens rarely), the tweet is labeled as \textit{neutral}.

When preprocessing tweets, we removed URLs because they normally do not represent relevant content but rather point to it. We also removed cash-tags (e.g., ``\$NKE'') and user mentions (e.g., ``@johndoe'') to make a tweet independent of a specific stock (company) and/or users involved in the discussion, and thus make the first step towards generalizing our model. Last but not least, we collapsed letter repetitions (e.g., ``coooool'' becomes ``cool''). This step is relatively easy to implement and has proven useful for sentiment classification tasks \cite{smailovic2015phd}. After these steps, we followed a typical bag-of-words computation procedure by applying tokenization (based on relatively simple regular expressions), lemmatization (we used LemmaGen \cite{Jursic10} for this purpose), $n$-gram construction (we included unigrams and bigrams into the feature set), and the TF-IDF weighting scheme \cite{Witten04}. Note that we did not remove stop words, such as ``not'', as this would in some cases change the sentiment polarity of a tweet.

\subsection*{Correlation and Granger causality}

For an initial investigation of the relation between the Twitter sentiment and stock prices, we apply the Pearson correlation and Granger causality tests. 
We use the Pearson correlation to measure the linear dependence between $P_d$ and  $R_d$.
Given two time series, $X_t$ and $Y_t$, the Pearson's correlation coefficient is calculated as:
\begin{equation}
\rho(X,Y)=\frac{\langle X_t Y_t\rangle-\langle X_t\rangle \langle Y_t\rangle}{\sqrt{(\langle X_t^2\rangle-\langle X_t\rangle^2)(\langle Y_t^2\rangle -\langle Y_t\rangle^2)}}
\end{equation}
where $\langle \cdot \rangle$ is the time average value.
The correlation $\rho(X,Y)$ quantifies the linear contemporaneous dependence. 

We also perform the Granger causality test \cite{granger1969investigating} to check if the Twitter variables help in the prediction of the price returns. 
The steps of the procedure applied are summarized as follows \cite{pivskorec2014cohesiveness}:
\begin{itemize}
\item Determine if the two time series are non-stationary, by the Augmented Dickey-Fuller (ADF) test.
\item Build a Vector Autoregressive (VAR) model and determine its optimal order by considering four measures: AIC, BIC, FPE, HQIC.
\item Fit the VAR model with the selected order from the previous step.
\item Perform the Ljung-box test for no autocorrelation in the residuals of the fit.
\item Perform the F-test to detect statistically significant differences in the fit of the baseline and the extended models (Granger causality test).
\end{itemize}

\subsection*{Event study}

The method used in this paper is based on 
an event study, as defined in financial econometrics \cite{mackinlay1997event}. 
This type of study analyzes the abnormal price returns observed during external events. 
It requires that a set of abnormal 
events for each stock is first identified (using prior knowledge or automatic detection), 
and then the events are grouped according to some measure of ``polarity'' 
(whether the event should have positive, negative or no effect on the 
valuation of the stock).
Then, the price returns for events of each group are analyzed.
In order to focus only on isolated events affecting a particular stock, the method removes the fluctuations (influences) of the market to which the stock belongs. This is achieved by using the market model, i.e., the price returns of a selected index.

\begin{figure}
\center
\includegraphics[width=0.55\textwidth]{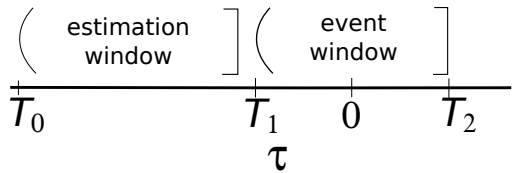}
 \caption{Time line for an event study.} 
 \label{graph}
\end{figure}

\paragraph*{Event window.} The initial task of conducting an event study is to define the events of interest and identify the period over which the stock prices of the companies involved in this event will be examined: the event window, as shown in Figure \ref{graph}. For example, if one is looking at the information content of an earnings announcement on day $d$, the event will be the earnings announcement and the event window $(T_1,T_2]$ might be $(d-1,d+1]$.
The reason for considering one day before and after the event is that the market may acquire information about the earnings prior to the actual announcement and one can investigate this possibility by examining pre-event returns. 



\paragraph*{Normal and abnormal returns.} To appraise the event's impact one needs a measure of the abnormal return. The abnormal return is the actual ex-post return of the stock over the event window minus the normal return of the stock over the event window. The normal return is defined as the return that would be expected if the event did not take place. For each company $i$ and event date $d$, we have:
\begin{equation}
AR_{i,d}=R_{i,d}- E[R_{i,d}]
\end{equation}
where $AR_{i,d}$, $R_{i,d}$, $E[R_{i,d}]$ are the abnormal, actual, 
and expected normal returns, respectively. 
There are two common choices for modeling the expected normal return: 
the constant-mean-return model, 
and the market model. 
The constant-mean-return model, as the name implies, assumes that the mean return of a given stock is constant through time. The market model, used in this paper, assumes a stable linear relation between the overall market return and the stock return.

\paragraph*{Estimation of the normal return model.} Once a normal return model has been selected, the parameters of the model must be estimated using a subset of the data known as the estimation window. The most common choice, when feasible, is to use the period prior to the event window for the estimation window (cf. Figure \ref{graph}). For example, in an event study using daily data and the market model, the market model parameters could be estimated over the 120 days prior to the event. Generally, the event period itself is not
included in the estimation period to prevent the event from influencing the normal 
return model parameter estimates.

\paragraph*{Statistical validation.} With the estimated parameters of the normal return model, the abnormal returns can be calculated. 
The null hypothesis, $H_0$, is that external events have no impact on the returns.
It has been shown that under $H_0$, abnormal returns are normally distributed, 
$AR_{i,\tau} \sim$ $\mathcal{N}(0,\sigma^2(AR_{i,\tau}))$ \cite{campbell1997econometrics}.
This forms the basis for a procedure which tests whether an abnormal 
return is statistically significant.


\paragraph*{Event detection using Twitter activity peaks.}

This part first discusses the algorithm used to detect Twitter activity peaks, which are then treated as events. Next, it describes the method used to assign a polarity to the events, using the Twitter sentiment. Finally, it discusses a specific type of events for the companies studied, called earnings announcement events, which are already known to produce abnormal price jumps.

\emph{Detection of Twitter peaks.} To identify Twitter activity peaks, for every company we use the time series of its daily Twitter volume, $TW_d$. We use a sliding window of $2L + 1$
days ($L = 5$) centered at day $d_0$,
and let $d_0$ slide along the time line. Within this window we evaluate the baseline volume
activity $TW_b$ as the median of the window \cite{lehmann2012dynamical}. 
Then, we define the outlier
fraction $\phi (d_0 )$ of the central time point $d_0$ as a relative difference of the activity $TW_{d_0}$ with respect to the median baseline
$TW_b$: $\phi (d_0)=[TW_d-TW_b]/max(  TW_b, n_{min} )$.
Here, $n_{min}=10$ is a minimum activity level used to regularize the definition 
of $\phi (d_0 )$ for low activity values. We say that there is an activity peak 
at $d_0$ if $\phi (d_0 ) > \phi _t$, where 
$\phi _t = 2$. 
The threshold $\phi _t$ determines the number of detected peaks and the overlaps
between the event windows --- both increase with larger $\phi _t$.
One should maximize the number of detected peaks, and minimize
the number of overlaps \cite{mackinlay1997event}.
We have analyzed the effects of varying $\phi _t$ from $0.5$ to $10$
(as in \cite{lehmann2012dynamical}). 
The decrease in the number of overlaps is substantial
for $\phi _t$ ranging from $0.5$ to $2$, for larger values the decrease is slower.
Therefore, we settled for $\phi _t = 2$.
As a final step we apply filtering which removes detected peaks
that are less then 21 days (the size of the event window) apart from the other peaks.

\begin{figure}
    \includegraphics[width=\textwidth]{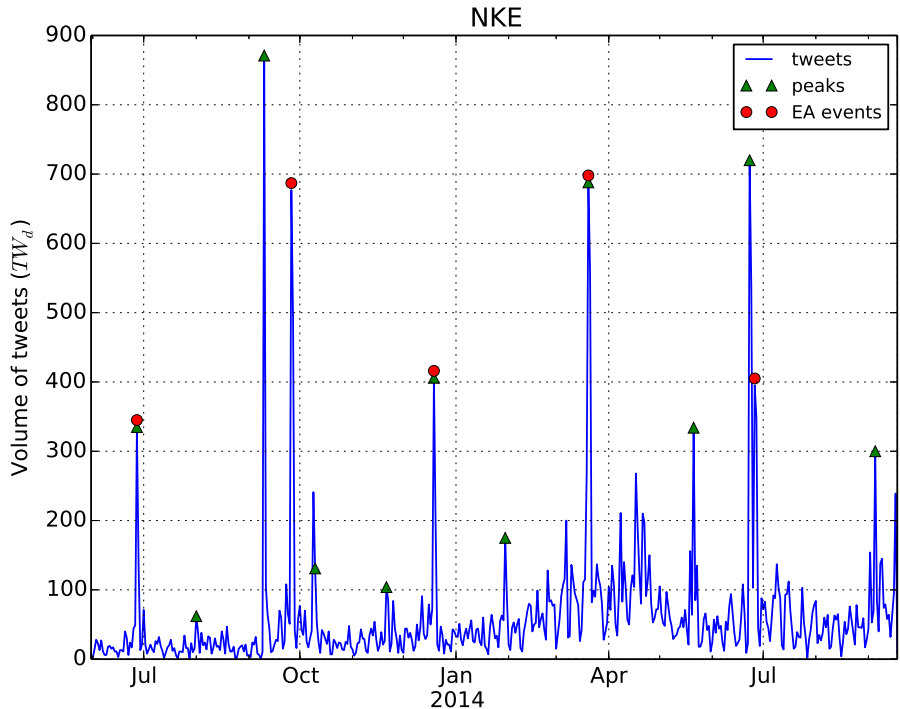}   
     \caption{Daily time series of Twitter volume with indicated peaks for the Nike company.} 
     \label{Nike}
\end{figure}

As an illustration, the resulting activity peaks for the Nike company are shown in Figure \ref{Nike}. After the peak detection procedure, we treat all the peaks detected as events.
These events are then assigned polarity (from Twitter sentiment) and type
(earnings announcement or not).

\begin{figure}
    \centering
	\includegraphics[width=0.95\textwidth]{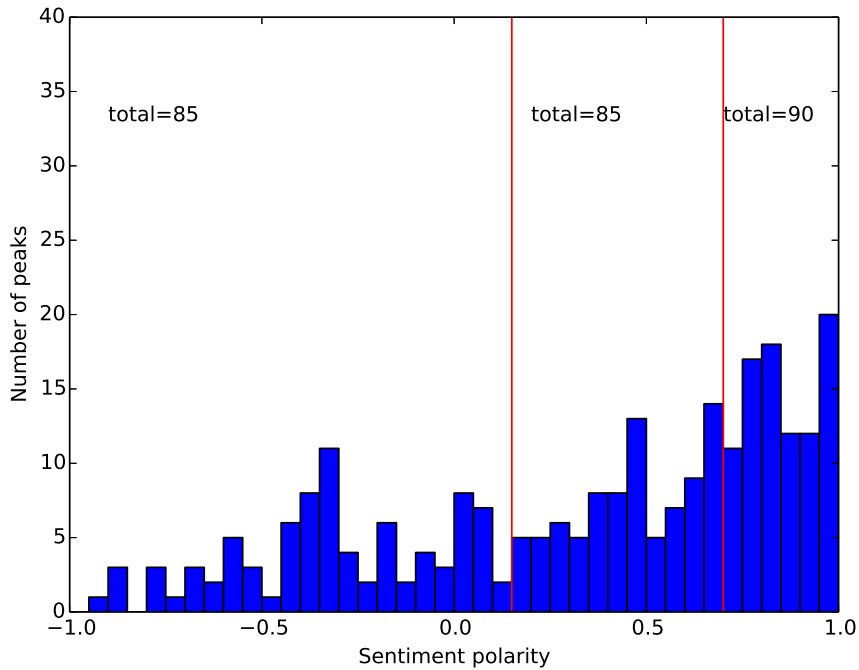}
 	\caption{Distribution of sentiment polarity for the 260 detected Twitter peaks. 
 	The two red bars indicate the chosen thresholds of the polarity values.}
    \label{peakspol4}
\end{figure}

\emph{Polarity of events.} 
Each event is assigned one of the three polarities: negative, neutral or positive. 
The polarity of an event is derived from the sentiment polarity $P_d$ 
of tweets for the peak day. 
From our data we detected 260 events.
The distribution of the $P_d$ values for the 260 events is not uniform, 
but prevailingly positive, as shown in Figure \ref{peakspol4}. 
To obtain three sets of events with approximately the same size,
we select the following thresholds, and define the event polarity as follows:
\begin{itemize}
\item If $P_d \in [-1, 0.15)$ the event is a \emph{negative event},
\item If $P_d \in [0.15, 0.7]$ the event is a \emph{neutral event},
\item If $P_d \in (0.7, 1]$ the event is a \emph{positive event}.
\end{itemize}
Putting thresholds on a signal is always somewhat arbitrary, 
and there is no systematic treatment of this issue in the event 
study \cite{mackinlay1997event}. 
The justification for our approach is that sentiment should be
regarded in relative terms, in the context of related events.
Sentiment polarity has no absolute meaning, but provides just an 
ordering of events on the scale from $-1$ (negative) to $+1$ (positive).
Then, the most straightforward choice is to distribute all the events
uniformly between the three classes.
Conceptually similar approaches, i.e., treating the sentiment in relative
terms, were already applied to compare the sentiment leaning of network communities towards
different environmental topics \cite{Sluban2015}, and to compare the
emotional reactions to conspiracy and science posts on Facebook
\cite{Zollo2015}.
Additionally, in the closely related work by Sprenger et al. 
\cite{Sprenger2014jbfa}, the authors use the percentage of positive tweets for 
a given day $d$, to determine the event polarity. 
Since they also report an excess of positive tweets,
they use the median share of positive tweets as a threshold between 
the positive and negative events.

\emph{Event types.} For a specific type of events in finance, in particular quarterly \emph{earnings announcements} (EA), it is known that the price return of a stock abnormally jumps in the direction of the earnings \cite{campbell1997econometrics,boehmer1991event}.
In our case, the Twitter data shows high posting activity during the EA events,
as expected. However, there are also other peaks in the Twitter activity, which do not
correspond to EA, abbreviated as non-EA events.
See Figure \ref{Nike} for an example of Nike.

The total number of peaks that our procedure detects in the period of the study is 260.
Manual examination reveals that in the same period, there are 151 EA events
(obtained from \url{http://www.zacks.com/}).
Our event detection procedure detects 118 of them, the rest are non-EA events.
This means that the recall (the fraction of all EA events that were correctly detected as EA)
of our peak detection procedure is 78\%.
In contrast, Sprenger et al. \cite{Sprenger2014jbfa} detect 224 out of 672
EA events, yielding the recall of 33\%. 
They apply a simpler peak detection procedure: a Twitter peak is defined
as one standard deviation increase of the tweet volume over the previous five days.

The number of the detected peaks 
indicates that there is a large number of interesting events on Twitter
which cannot be explained by earnings announcement. 
The impact of the EA events on price returns is already known in the literature,
and our goal is to reconfirm these results.
On the other hand, the impact of the non-EA events is not known, and
it is interesting to verify if they have similar impact on prices as the EA events.

Therefore, we perform the event study in two scenarios, with explicit detection of the two
types of events, all the events (including EA) and non-EA events only:
\begin{enumerate}
\item Detecting {\bf all events} from the complete time interval of the data, 
including the EA days.
In total, 260 events are detected, 118 out of these are the EA events.
\item Detecting {\bf non-EA events} from a subset of the data.
For each of the 151 EA events, where $d$ is the event day, we first remove the interval $[d-1,d+1]$, 
and then perform the event detection again. This results in 182 non-EA events detected.  
\end{enumerate}

We report all the detected peaks, for the EA and non-EA events,
with the dates and their polarity, in \nameref{S1_Text}.


The first scenario allows to compare the results of the Twitter sentiment  
with the existing literature in financial econometrics \cite{campbell1997econometrics}. 
It is worth noting, however, that 
the variable used to infer ``polarity'' of the events there is the
difference between the expected and announced earnings.
The analysis of the non-EA events in the second scenario tests 
if the Twitter sentiment data contains useful information about 
the behavior of investors for other types of events, 
in addition to the already well-known EA events.


\paragraph*{Estimation of normal returns.}
Here we briefly explain the market model procedure for estimation of normal returns.
Our methodology follows the one presented in \cite{campbell1997econometrics} and \cite{malkiel1970efficient}.
The market model is a statistical model which relates the return of a
given stock to the return of the market portfolio. 
The model's linear specification follows from the assumed joint
normality of stock returns.
We use the DJIA index as a normal market model.
This choice helps us avoid adding too many variables to our model and simplifies the computation of the result.
The aggregated DJIA index is computed from the mean 
weighted prices of all the stocks in the index. 
For any stock $i$, and date $d$, the market model is:
\begin{eqnarray}
R_{i,d}&=&  \alpha_i + \beta_i R_{DJIA,d} +\epsilon_ {i,d}\\
E(\epsilon_ {i,d})&=&0,\;\; var(\epsilon_ {i,d}) = \sigma_{\epsilon_ {i,d}}^2 \\
E[R_{i,d}] &=& \hat{\alpha_i} + \hat{\beta_i} R_{DJIA,d}
\end{eqnarray}
where $R_{i,d}$ and $R_{DJIA,d}$ are the returns of stock $i$ and the market portfolio, respectively, and  $\epsilon_ {i,d}$ is the zero mean disturbance term.
$\alpha_i , \beta_i, \sigma_{\epsilon_ {i,d}}^2$
are the parameters of the market model.
To estimate these parameters for a given event and stock, we use an estimation window of $L=120$ days,
according to the hint provided in \cite{campbell1997econometrics}.
Using the notation presented in Figure \ref{graph} for the time line, the estimated value of $\sigma_{\epsilon_ {i,d}}^2$ is:
\begin{equation}
\hat{\sigma}_{\epsilon_{i,d}}^2=\frac{1}{L-2}\sum_{d=T_0+1}^{T_1}(R_{i,d}- \hat{\alpha_i} - \hat{\beta_i} R_{DJIA,d})^2
\end{equation}
where $\hat{\alpha_i},\hat{\beta_i}$ are the estimated parameters following the OLS procedure\cite{campbell1997econometrics}.
The abnormal return for company $i$ at day $d$ is the residual :
\begin{equation}
AR_{i,d}=R_{i,d}-  \hat{\alpha_i} - \hat{\beta_i} R_{DJIA,d}.
\end{equation}

\paragraph*{Statistical validation.}
Our null hypothesis, $H_0$, is that
external events have no impact on the behavior of returns (mean or variance).
The distributional properties of the abnormal returns can be used to draw
inferences over any period within the event window. 
Under $H_0$, the distribution of the sample abnormal return of a given observation 
in the event window is normal:
\begin{equation}
AR_{i,\tau} \thicksim \mathcal{N}(0, \sigma_{AR}^2) \; \textcolor{blue}{.}
\label{ho}
\end{equation}
Equation \ref{ho} takes into account the aggregation of the abnormal returns. 

The abnormal return observations must be aggregated in order to draw
overall conclusions for the events of interest. 
The aggregation is along two dimensions: through time and across stocks. 
By aggregating across all the stocks \cite{malkiel1970efficient}, we get:
\begin{equation}						
\overline{AR}_\tau=(1/N)\sum_{i=1}^{N}AR_{i,\tau} \; \textcolor{blue}{.}
\end{equation}
The cumulative abnormal return ($CAR$) from time $\tau_1$ to $\tau_2$ is the sum of the abnormal returns: 
\begin{equation}
CAR(\tau_1,\tau_2) = \sum_{\tau=\tau_1}^{\tau_2}\overline{AR}_\tau \; \textcolor{blue}{.}
\end{equation}
To calculate the variance of the $CAR$, we assume $\sigma_{AR}^2=\sigma_{\epsilon_ {i,t}}^2$  (shown in e.g., \cite{campbell1997econometrics,malkiel1970efficient}):
\begin{equation}
var(CAR(\tau_1,\tau_2))= (1/N^2)\sum_{i=1}^{N}(\tau_2 - \tau_1 +1)\sigma_{\epsilon_i}^2 
\end{equation}
where $N$ is the total number of events.
Finally, we introduce the test statistic $\hat{\theta}$. With this quantity we can test if the measured return is abnormal:
\begin{equation}
\frac{CAR(\tau_1,\tau_2)}{\sqrt[2]{var(CAR(\tau_1,\tau_2))}}=  \hat{\theta} \thicksim  \mathcal{N}(0,1) 
\end{equation}
where $\tau$ is the time index inside the event window, and $|\tau_2 - \tau_1|$   
is the total length of the event window. 

\section*{Results}
This section first presents an exhaustive evaluation of the Twitter sentiment classification model.
Then it shows the correlation and Granger causality results over the entire time period.
Finally, it shows statistically significant results of the event study methodology as
applied to Twitter data.

\subsection*{Twitter sentiment classification}

In machine learning, a standard approach to evaluate a classifier is by cross-validation.
We have performed a 10-fold cross-validation on the set of 103,262 annotated tweets.
The whole training set is randomly partitioned into 10 folds, one is set apart for testing,
and the remaining nine are used to train the model and evaluate it on the test fold.
The process is repeated 10 times until each fold is used for testing exactly once.
The results are averaged over 10 tests and from standard deviations the 95\% confidence
intervals are computed. The results are given in Table~\ref{tbl_agree_class}.

Cross-validation gives an estimate of the sentiment classifier performance on the application
data, assuming that the training set is representative of the application set.
However, it does not provide any hint about the highest performance achievable.
We claim that the agreement between the human experts
provides an upper bound that the best automated classifier can achieve.
The inter-annotator agreement is computed from a fraction of tweets annotated twice.
During the annotation process, 6,143 tweets were annotated twice, by two different annotators.
The results were used to compute various agreement measures.

There are several measures to evaluate the 
performance of classifiers and compute the inter-annotator agreement.
We have selected the following three measures 
to estimate and compare them:
\acc, \accone, and $\overline{F_{1}}$.
\acc($-,0,+$) is the fraction of correctly classified
examples for all three sentiment classes. This is the simplest and most common
measure, but it doesn't take into account the ordering of the classes.
On the other extreme, \accone($-,+$) (a shorthand for \textit{Accuracy within 1} neighboring class)
completely ignores the neutral class.
It counts as errors just the negative sentiment examples predicted as
positive, and vice versa.
$\overline{F_{1}}(-,+)$ is the average of $F_{1}$ for the negative and 
positive class.
It does not account for the misclassification of the neutral class
since it is considered less important than the extremes, i.e., negative or positive sentiment.
However, the misclassification of the neutral sentiment is taken into account 
implicitly as it affects the precision and recall of the extreme classes.
$F_{1}$ is the harmonic mean of \prc\, and \rcl\, for each class.
\prc\, is a fraction of correctly predicted examples out of all the predictions of a particular class.
\rcl\, is a fraction of correctly predicted examples out of all actual members of the class.
$\overline{F_{1}}(-,+)$ is a standard measure of performance for sentiment classifiers\cite{kiritchenko2014sentiment}.

\begin{table}[h]
	\begin{tabular}{|l|c|c|}
		\hline\bf {  }  & \bf {Annotator agreement} & \bf {Sentiment classifier}  \\ 
		\hline
		No. of hand-labeled examples & $ 6,143 $   & $ 103,262 $\\ \hline
		\acc($-,0,+$)           & $ 77.1\% $    & $ 76.0\pm\!0.5\% $ \\ \hline
		\accone($-,+$)          & $ 98.8\% $    & $ 99.4\pm\!0.1\% $ \\ \hline
		$\overline{F_{1}}(-,+)$ & $ 49.4\% $    & $ 50.8\pm\!1.0\% $ \\ \hline
	    \hspace*{1em}\prc/\rcl($-$)          & $ 48.0/48.0\% $    & $ 71.3/38.9\% $ \\ \hline
	    \hspace*{1em}\prc/\rcl($+$)          & $ 50.9/50.9\% $    & $ 68.6/40.9\% $ \\ \hline
	\end{tabular}

\caption{The inter-annotator agreement (on the examples labeled twice) and the classifier
        performance (from 10-fold cross-validation) over several evaluation measures.}
\label{tbl_agree_class}
\end{table}

Table~\ref{tbl_agree_class} gives the comparison of the inter-annotator agreement and
the classifier performance. 
The classifier has reached the annotator agreement in all three measures.
In a closely related work by Sprenger et al. \cite{Sprenger2014jbfa},
they use Naive Bayes for sentiment classification. Their classifier is
trained on 2,500 examples, and the 10-fold cross-validation yields \acc\,
of 64.2\%.

We argue that in our case, there is no need to invest further work to improve the classifier.
Most of the hypothetical improvements would likely be the result of
overfitting the training data. We speculate that
the high quality of the sentiment classifier is mainly the consequence of a large
number of training examples. In our experience in diverse domains, one needs
about 50,000 -- 100,000 labeled examples to reach the inter-annotator agreement.

If we compare the $F_{1}$ measures, we observe a difference in the respective
\prc\, and \rcl. For both classes, $-$ and $+$, the sentiment classifier has 
a considerably higher \prc, at the expense of a lower \rcl.
This means that tweets, classified into extreme sentiment classes ($-$ or $+$) are
likely indeed negative or positive (\prc\, about 70\%), even if the classifier 
finds only a smaller fraction of them (\rcl\, about 40\%).
This suits well the purpose of this study.
Note that it is relatively easy to modify the SVM classifier, without retraining it,
to narrow the space of the neutral class, thus increasing the recall of
the negative and positive classes, and decreasing their precision.
One possible criterion for such a modification is to match
the distribution of classes in the application set, as predicted
by the classifier, to the actual distribution
in the training set.

\subsection*{Correlation and Granger causality}
\emph{Correlation.} Table \ref{GrangerResults} shows the computed Pearson 
correlations, as defined in the Methods section.
The computed coefficients are small, but are in line with the result of \cite{mao2011predicting}. 
In our opinion, these findings and the one published in \cite{mao2011predicting} underline that when considering the entire time period of the analysis, days with a low number of tweets affect the measure. 

\begin{table}
    \centering
    \begin{tabular}{l|c|lcc|cc}
    \hline
     Ticker & Pearson correlation      & {} & \multicolumn{4}{c}{ Granger causality } \\ \hline
           & $\rho(P_d, R_d)$      & {}& \multicolumn{2}{c|}{ $P_d$ \& $R_d$ } & \multicolumn{2}{c}{ $TW_d$ \& $|R_d|$} \\ \hline
     TRV   & 0.1178  &        &  ←   & ~   & ~   & ~      \\
     UNH   & 0.2565  &        &  ←   &     &     &        \\
     UTX   & 0.1370  &        &  ←   &     &     &        \\
     MMM   & 0.1426  &        &  ←   & ~   &  ←  &        \\
     DD    & 0.2680  &        &  ←   &     &     &        \\
     AXP   & 0.1566  &        &  ←   & ~   &     &  →     \\
     PG    & 0.2145  &        &      &     &     &        \\
     NKE   & 0.2460  &        &      &     &     &        \\
     CVX   & 0.2053  &        &      &     &     &        \\
     HD    & 0.2968  &        &  ←   &     &     &  →     \\
     CAT   & 0.3648  &        &      &     &     &        \\
     JNJ   & 0.2220  &        &      &     &     &        \\
     V     & 0.2995  &        &  ←   &     &     &        \\
     VZ    & 0.1775  &        &      &     &     &        \\
     KO    & 0.1203  &        &      &     &     &        \\
     MCD   & 0.2047  &        &      &     &     &  →     \\
     XOM   & 0.2738  &        &      &     &  ←  &        \\
     DIS   & 0.2305  &        &  ←   &     &     &  →     \\
     BA    & 0.2408  &        &      &     &     &  →     \\
     MRK   & 0.1758  &        &      &     &     &        \\
     CSCO  & 0.2393  &        &      &  →  &     &  →     \\
     GE    & 0.1450  &        &      &     &     &        \\
     WMT   & 0.2710  &        &      &  →  &     &        \\
     INTC  & 0.2703  &        &      &     &     &  →     \\
     PFE   & 0.1252  &        &      &     &     &        \\
     T     & 0.1409  &        &      &     &     &  →     \\
     GS    & 0.3405  &        &      &     &     &        \\
     IBM   & 0.3462  &        &      &  →  &     &  →     \\
     JPM   & 0.1656  &        &  ←   &     &     &        \\
    MSFT   & 0.2700  &        &      &     &     &  →     \\ 
    \hline
    ~      &   ~          & ~      & 10   & 3   & 2   & 10     \\ 
    \hline
    \end{tabular}

    \caption{Results of the Pearson correlation and Granger causality tests. 
    Companies are ordered as in Table \ref{tbl_stocks}.
    The arrows indicate a statistically significant Granger causality relation 
    for a company, at the 5\% significance level. 
    A right arrow indicates that the Twitter variable 
    (sentiment polarity $P_d$ or volume $TW_d$) 
    Granger-causes the market variable (return $R_d$), while a left arrow indicates that the 
    market variable Granger-causes the Twitter variable. 
    The counts at the bottom show the total number of companies passing the Granger test.} 
     \label{GrangerResults}
\end{table}

\emph{Granger causality.}   
The results of the Granger causality tests are also in Table \ref{GrangerResults}. 
They show the results of the causality test in both directions: 
from the Twitter variables to the market variables and vice versa. 
The table gives the Granger causality links per company between 
a) sentiment polarity and price return, and 
b) the volume of tweets and absolute price return. 
The conclusions that can be drawn are:
\begin{itemize}
\item The polarity variable is not useful for predicting the price return, 
as only three companies pass the Granger test. 
\item The number of tweets for a company Granger-causes the absolute price return for one third of the companies. This indicates that the amount of attention on Twitter is useful for predicting the price volatility. Previously, this was known only for an aggregated index, 
but not for individual stocks \cite{mao2011predicting,bollen2011twitter}.
\end{itemize}

\subsection*{Cumulative abnormal returns}
\begin{figure}[ht]
	\includegraphics[width=\textwidth]{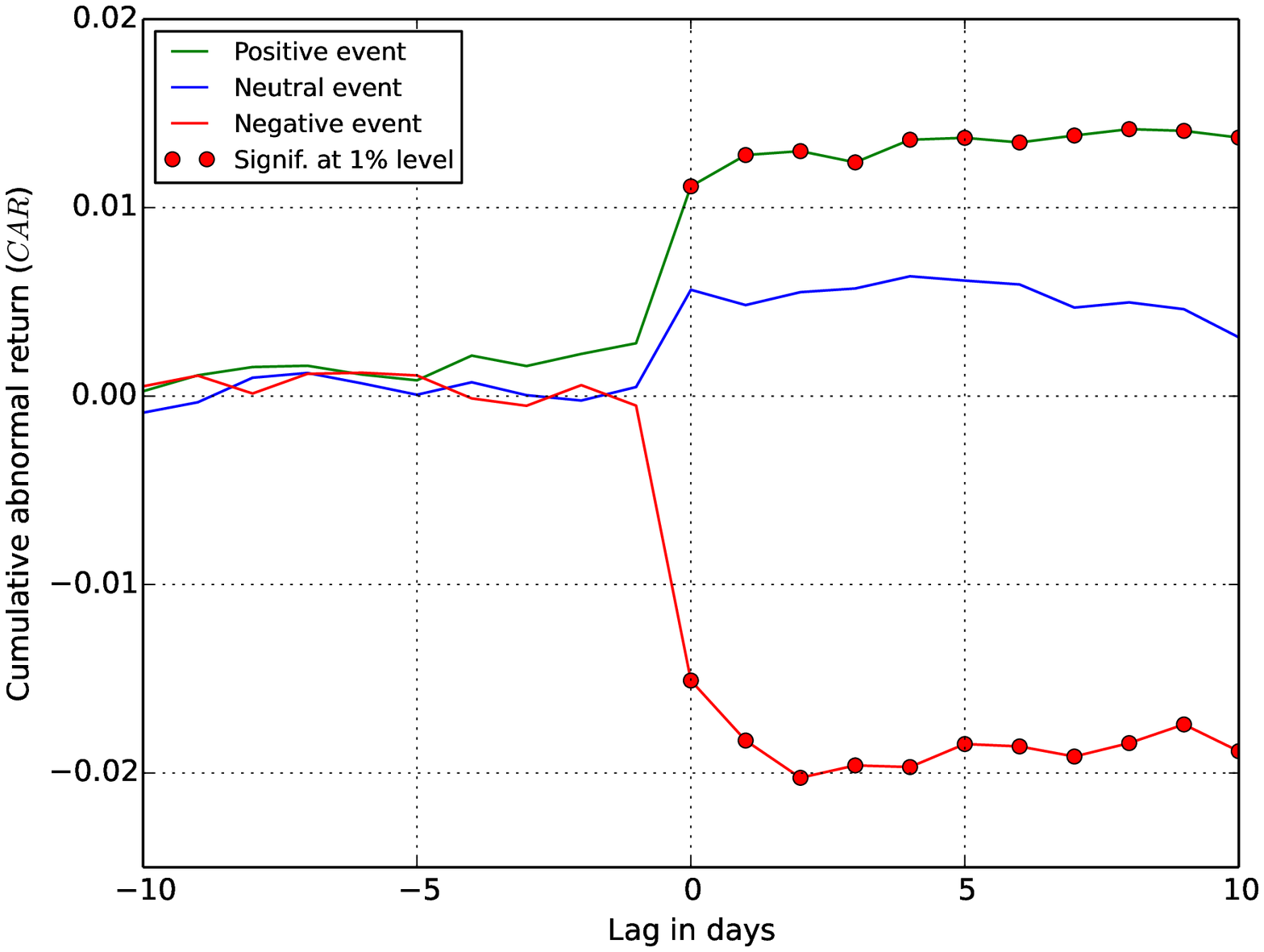}
 	\caption{$CAR$ for all detected events, including EA.
    The $x$ axis is the lag between the event and $CAR$, 
    and the red markers indicate days with statistically 
    significant abnormal return.}
 	\label{eapeaks4}
\end{figure}

\begin{figure}[ht]
	\includegraphics[width=\textwidth]{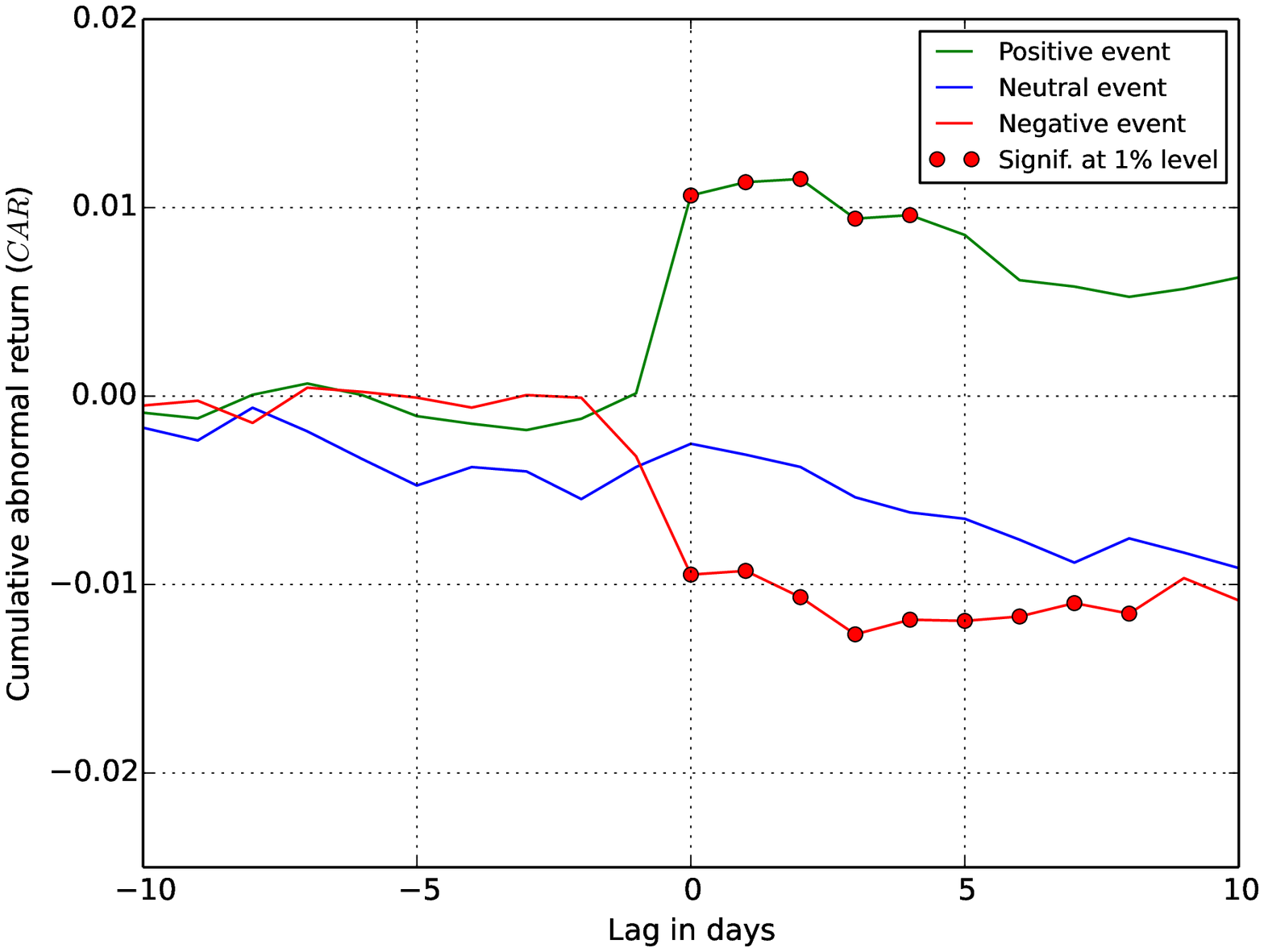}
 	\caption{$CAR$ for non-EA events. The $x$ axis is the lag between the event and $CAR$, 
 	and the red markers indicate days with statistically significant abnormal return.}
 	\label{twpeaks4}
\end{figure}

The results of the event study are shown in Figures \ref{eapeaks4} and \ref{twpeaks4}, 
where the cumulative abnormal returns ($CAR$) are plotted for the two types of events defined earlier.
The results are largely consistent with the existing literature on
the information content of earnings \cite{campbell1997econometrics,boehmer1991event}.
The evidence strongly supports the hypothesis that tweets do indeed convey information
relevant for stock returns.

Figure \ref{eapeaks4} shows $CAR$ for all the detected Twitter peaks, including 
the EA events (45\% of the detected events are earnings announcements).
The average $CAR$ for the events is abnormally increasing after the positive peaks 
and decreasing after the negative sentiment peaks.
This is confirmed with details in Table \ref{thetas}.
The values of $CAR$ are significant at the 1\% level
for ten days after the positive sentiment events.
Given this result, the null hypothesis that the
events have no impact on price returns is rejected.
The same holds for negative sentiment events, but
the $CAR$ (actually loss) is twice as large in absolute terms.
The $CAR$ after the neutral events is very low,
and barely significant at the 5\% level at two days;
at other days one cannot reject the null hypothesis.
We speculate that the positive $CAR$ values for the neutral events, 
barely significant, are the result of the uniform distribution of 
the Twitter peaks into three event classes (see Figure \ref{peakspol4}).
An improvement over this baseline approach remains a subject of
further research.

A more interesting result concerns the non-EA events in Figure \ref{twpeaks4}.
Even after removing the earnings announcements, with already known
impact on price returns, one can reject the null hypothesis.
In this case, the average $CAR$ of the non-EA
events is abnormally increasing after the detected positive peaks and
decreasing after the negative peaks.
Table \ref{thetas} shows that after the event days the values of
$CAR$ remain significant at the 1\% level for four
days after the positive events,
and for eight days after the negative events.
The period of impact of Twitter sentiment on price returns is
shorter when the EA events are removed, and the values of $CAR$ are lower,
but in both cases the impact is statistically significant.
The $CAR$ for the neutral events tend to be slightly negative (in contrast to the
EA events), albeit are not statistically significant. However, this again
indicates that the distribution of Twitter peaks into the event classes 
could be improved.

\begin{table}[ht]
\caption{Values of the $\hat{\theta}$ statistic 
for each type of event. Significant results at the 1\% significance level ($|\hat{\theta}|>2.58$) are denoted by **, and at the 5\% level ($|\hat{\theta}| > 1.96$) by *. }
\label{thetas}
\begin{tabular}{crrrrrr}
\hline
Lag & \multicolumn{3}{c}{All events (including EA)} & \multicolumn{3}{c}{Non-EA events} \\
(days) &   negative &    neutral &   positive &   negative &    neutral &   positive \\
\hline

-10 &  0.6408\sn & -1.0730\sn &  0.3208\sn & -0.5281\sn & -1.5168\sn & -1.0017\sn \\
-9  &  0.9495\sn & -0.2828\sn &  0.9806\sn & -0.1847\sn & -1.5060\sn & -0.9509\sn \\
-8  &  0.0977\sn &  0.6852\sn &  1.1197\sn & -0.8646\sn & -0.3225\sn &  0.0458\sn \\
-7  &  0.7302\sn &  0.7470\sn &  1.0126\sn &  0.2333\sn & -0.8464\sn &  0.3790\sn \\
-6  &  0.6865\sn &  0.3657\sn &  0.6419\sn &  0.1069\sn & -1.3505\sn &  0.0276\sn \\
-5  &  0.5536\sn &  0.0356\sn &  0.4295\sn & -0.0358\sn & -1.7525\sn & -0.4941\sn \\
-4  & -0.0580\sn &  0.3377\sn &  1.0212\sn & -0.2430\sn & -1.2873\sn & -0.6304\sn \\
-3  & -0.2255\sn &  0.0207\sn &  0.7089\sn &  0.0200\sn & -1.2781\sn & -0.7248\sn \\
-2  &  0.2395\sn & -0.0961\sn &  0.9382\sn & -0.0302\sn & -1.6476\sn & -0.4560\sn \\
-1  & -0.1981\sn &  0.1849\sn &  1.1148\sn & -1.0632\sn & -1.0765\sn &  0.0535\sn \\
 0  & -5.6350\sd &  2.0709\se &  4.2197\sd & -3.0057\sd & -0.6897\sn &  3.6489\sd \\
 1  & -6.5332\sd &  1.6975\sn &  4.6436\sd & -2.8173\sd & -0.8118\sn &  3.7254\sd \\
 2  & -6.9559\sd &  1.8629\sn &  4.5338\sd & -3.1146\sd & -0.9436\sn &  3.6325\sd \\
 3  & -6.4855\sd &  1.8582\sn &  4.1682\sd & -3.5557\sd & -1.2979\sn &  2.8611\sd \\
 4  & -6.2936\sd &  1.9989\se &  4.4168\sd & -3.2240\sd & -1.4419\sn &  2.8187\sd \\
 5  & -5.7154\sd &  1.8655\sn &  4.3086\sd & -3.1383\sd & -1.4721\sn &  2.4297\se \\
 6  & -5.5829\sd &  1.7492\sn &  4.1047\sd & -2.9850\sd & -1.6720\sn &  1.6956\sn \\
 7  & -5.5822\sd &  1.3478\sn &  4.0987\sd & -2.7250\sd & -1.8837\sn &  1.5573\sn \\
 8  & -5.2308\sd &  1.3889\sn &  4.0868\sd & -2.7867\sd & -1.5667\sn &  1.3732\sn \\
 9  & -4.8243\sd &  1.2552\sn &  3.9575\sd & -2.2729\se & -1.6803\sn &  1.4462\sn \\
 10 & -5.0916\sd &  0.8288\sn &  3.7645\sd & -2.4901\se & -1.8009\sn &  1.5622\sn \\

\hline
\end{tabular}
\end{table}

These results are similar to the ones reported by Sprenger at al. \cite{Sprenger2014jbfa}.
In addition, the authors show statistically significant 
increase in the $CAR$ values even before the positive event days.
They argue that this is due to the information leakage before the
earnings announcements.
We observe a similar phenomena, but with very low $CAR$ values, and not 
statistically significant
(cf. the positive events at day $-1$ in Figure \ref{eapeaks4}).

\section*{Discussion}
In this work we present significant evidence of dependence between
stock price returns and Twitter sentiment in tweets about the
 companies. 
As a series of other papers have already shown, there is a signal 
worth investigating which connects social media and market behavior.
This opens the way, if not to forecasting, then at least 
to ``now-casting'' financial markets. 
The caveat is that this dependence becomes useful only when data are properly selected, or different sources of data are analyzed together.
For this reason, in this paper, we first identify events, marked
by increased activity of Twitter users, and then observe market
behavior in the days following the events.
This choice is due to our hypothesis that only at some moments,
identified as events, there is a strong interaction between 
the financial market and Twitter sentiment.
Our main result is that the aggregate Twitter sentiment 
during the events implies the direction of market evolution.
While this can be expected for peaks related to ``known'' events, like earnings announcements, it is really interesting to note that a similar conclusion holds also when peaks do not correspond to any expected 
news about the stock traded.

Similar results were corroborated in a recent, independent study by 
Sprenger et al. \cite{Sprenger2014jbfa}.
The authors have made an additional step, and classified the non-EA events
into a comprehensive set of 16 company-specific categories.
They have used the same training set of 2,500 manually classified tweets
to train a Naive Bayesian classifier which can then reasonably well discriminate
between the 16 categories.
In our future work, we plan to identify topics, which are not predefined,
from all the tweets of non-EA events. 
We intend to apply standard topic detection algorithms, such as Latent Dirichlet
allocation (LDA) or clustering techniques.

Studies as this one could be well used in order to establish a direct relation 
between social networks and market behavior. 
A specific application could, for example, detect and possibly 
mitigate panic diffusion in the market from social network analysis.
To such purpose there is some additional research to be done in
the future. 
One possible direction is to test the presence of forecasting power 
of the sentiment time series. Following an approach similar to the one 
presented by Moat et al. \cite{moat2013quantifying} one can decide 
to buy or sell according to the presence of a peak in the tweet volume 
and the level of polarity in the corresponding direction.
However, detection of Twitter events should rely just on
the current and past Twitter volume.

Also, during the events, we might move to a finer time scale,
e.g., from daily to hourly resolution, as done by \cite{Zheludev2014}.
Finally, our short term plan is to extend the analysis to
a larger number of companies with high Twitter volume,
and over longer period of time.

\section*{Acknowledgments}

The authors would like to thank Sowa Labs
for providing the raw Twitter data 
and the annotated tweets of the DJIA 30 stocks for this research,
and to Sa\v{s}o Rutar for running the annotator agreement and
sentiment classification evaluations.


\section*{Supporting Information}

\label{S1_Text} {\bf S1 Appendix. Event dates and polarity.} Detailed information about the detected events from the Twitter data and their polarity. We show the 118 detected EA events and 182 detected non-EA events. 

\end{document}